\documentclass[acmsmall,nonacm]{acmart}

\setcopyright{none}
\renewcommand\footnotetextcopyrightpermission[1]{}

\usepackage{booktabs}
\usepackage{listings}
\usepackage{array}
\usepackage{tikz}
\usetikzlibrary{arrows.meta,calc,positioning,shapes.multipart}

\newcommand{\Rocq}{Rocq}
\newcommand{\Coq}{Coq}
\newcommand{\CertiGC}{CertiGC}
\newcommand{\CertiGraph}{CertiGraph}
\newcommand{\VST}{VST}
\newcommand{\tus}{\_\allowbreak{}}
\newcolumntype{L}[1]{>{\raggedright\arraybackslash}p{#1}}

\title[AI-Assisted Completion of CertiGC Proofs]{AI-Assisted Completion of
\CertiGC{} Proofs:\texorpdfstring{\\ }{ }An Experience Report}

\author{Shengyi Wang}
\affiliation{%
  \institution{Shanghai Qi Zhi Institute}
  \city{Shanghai}
  \country{China}
}
\email{wangshengyi@sqz.ac.cn}

\begin{abstract}
This experience report describes the Codex-assisted completion and stabilization
of a substantial \Rocq{} (formerly \Coq{}) proof development for \CertiGC{}, the
verified generational garbage collector in the \CertiGraph{} project.  The
development extends the collector from an effectively immutable setting to a
mutable one by adding remembered-set forwarding to the collection path and
re-establishing the top-level graph-isomorphism correctness theorem.  The
central technical issue was not low-level proof scripting alone: mutable updates
invalidate the old global no-backward-edge assumption, so the proof had to be
reorganized around a recorded-backward-edge invariant stating that every
backward edge is recorded in the appropriate remembered-set component.

This case differs from recent AI-assisted formal-proof accounts: unlike the
adaptation of a nearby compiler-proof architecture or a fresh metatheory
formalization, it completes a long-running verification in a mature codebase
built on the Verified Software Toolchain (\VST{}) and \CertiGraph{}, both
mechanized in \Rocq{}.
The \Rocq{} kernel remained the arbiter of correctness, while our role
shifted toward adjudicating invariant proposals, constraining specification
changes, reviewing theorem statements, and deciding when proof cleanup was
justified.
The Codex-assisted phase repaired the \VST{} relation proofs first, then
restored the mathematical graph-isomorphism theorem, and only then audited the
premise path from the \VST{} specification to the theorem.  That audit found and
removed a stale no-backward-edge condition from the \VST{}-facing proof path.
This report presents the workflow, resulting proof artifact, and lessons for
agentic proof maintenance.
\end{abstract}

\begin{document}

\maketitle

\section{Introduction}

Large language models are now routinely used as programming
assistants, and proof engineering is beginning to adopt the same style
of assistance.  In ordinary software development, compilation is only
a weak check on behavior.  In formal verification, by contrast, a
proof script accepted by the kernel establishes its stated theorem.
The central risk is therefore not that one must trust the generated
proof script, but that the definitions, theorem statement, or
specification boundary may no longer express the intended result.  A
silently weakened theorem, an inappropriate new precondition, or a
specification change can make a checked proof certify the wrong
property.  This does not make review automatic, but it changes its
scale: the theorem statements, together with the definitions and
specification boundaries they mention, are far more concise than the
proof scripts themselves.

This paper reports on a concrete proof-engineering experience in that
setting: using OpenAI Codex \citep{openai2025codex,openai2026codexapp}
to help complete the mutable-garbage-collector branch of \CertiGC{}.
\CertiGC{} is part of the \CertiGraph{} development
\citep{wang2019certigraph}, which uses the Verified Software Toolchain
(\VST{}) \citep{appel2014programlogics} and \Rocq{} to verify C
programs manipulating heap-represented graphs.  \Rocq{} is the proof
assistant formerly named \Coq{}.  The collector is a generational
copying collector.  The original CertiGC proof did not cover mutable
references or updatable arrays, whose updates can create pointers from
older objects to younger ones after allocation.  The mutable extension
adds a write barrier and remembered sets, giving the collector the
extra information needed to handle such updates.

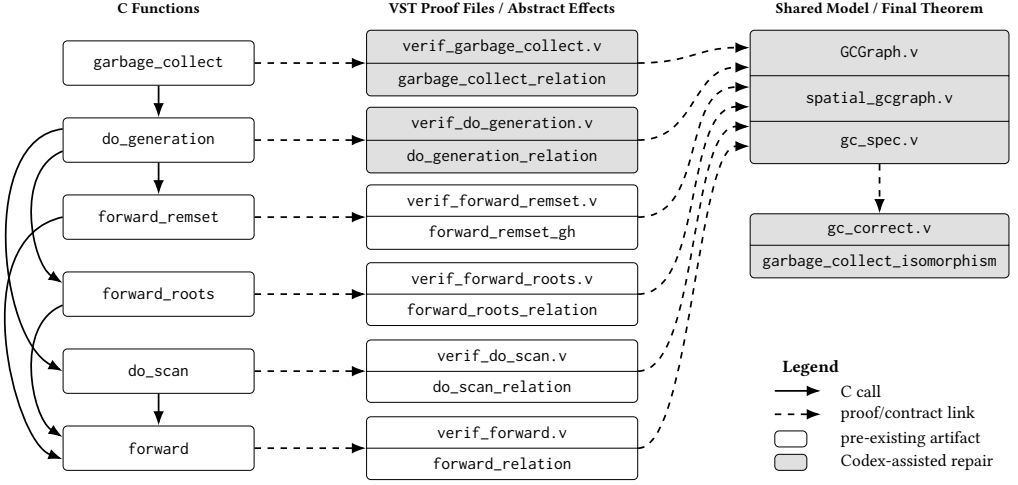
\begin{figure}[htbp]
\centering
\resizebox{\textwidth}{!}{\begin{tikzpicture}[
  font=\footnotesize,
  >=Latex,
  box/.style={
    draw,
    rounded corners=2pt,
    align=center,
    inner xsep=5pt,
    inner ysep=4pt,
    minimum height=0.72cm
  },
  cbox/.style={box, minimum width=3.1cm},
  smallc/.style={box, minimum width=3.1cm},
  widec/.style={box, minimum width=3.1cm},
  splitbox/.style={
    box,
    shape=rectangle split,
    rectangle split parts=2,
    rectangle split part align={center,center}
  },
  proofbox/.style={splitbox, minimum width=4.4cm, minimum height=1.06cm},
  theorem/.style={splitbox, minimum width=4.2cm, minimum height=1.45cm},
  specbox/.style={
    box,
    shape=rectangle split,
    rectangle split parts=3,
    rectangle split part align={center,center,center},
    minimum width=4.2cm,
    inner ysep=7pt
  },
  existing/.style={fill=white},
  repaired/.style={fill=black!12},
  call/.style={->, thick, rounded corners=3pt},
  prooflink/.style={->, dashed, thick},
  note/.style={align=center, font=\scriptsize},
  legendtext/.style={font=\footnotesize, anchor=west}
]

\node[note] at (0,0.45) {\textbf{C Functions}};
\node[note] at (5.575,0.45) {\textbf{VST Proof Files / Abstract Effects}};
\node[note] at (11.7,0.45) {\textbf{Shared Model / Final Theorem}};

\node[cbox,existing] (gc) at (0,-0.45) {\texttt{garbage\_collect}};
\node[cbox,existing] (dg) at (0,-1.70) {\texttt{do\_generation}};
\node[smallc,existing] (frs) at (0,-2.95) {\texttt{forward\_remset}};
\node[smallc,existing] (fr) at (0,-4.20) {\texttt{forward\_roots}};
\node[smallc,existing] (ds) at (0,-5.45) {\texttt{do\_scan}};
\node[widec,existing] (fwd) at (0,-6.70) {\texttt{forward}};

\draw[call] (gc) -- (dg);
\draw[call] (dg) -- (frs);
\draw[call] ($(dg.west)+(0,-0.18)$)
  .. controls +(-0.80,-0.32) and +(-0.50,0.32) ..
  ($(fr.west)+(0,0.18)$);
\draw[call] ($(dg.west)+(0,0.18)$)
  .. controls +(-1.35,-0.15) and +(-1.05,0.55) ..
  (ds.west);
\draw[call] (frs.west)
  .. controls +(-1.60,-0.35) and +(-0.80,0.30) ..
  ($(fwd.west)+(0,-0.18)$);
\draw[call] ($(fr.west)+(0,-0.18)$)
  .. controls +(-0.80,-0.32) and +(-0.50,0.32) ..
  ($(fwd.west)+(0,0.18)$);
\draw[call] (ds) -- (fwd);

\node[proofbox,repaired] (vgc) at (5.575,-0.45)
  {\texttt{verif\_garbage\_collect.v}
   \nodepart{two}
   \texttt{garbage\_collect\_relation}};
\node[proofbox,repaired] (vdg) at (5.575,-1.70)
  {\texttt{verif\_do\_generation.v}
   \nodepart{two}
   \texttt{do\_generation\_relation}};
\node[proofbox,existing] (vrs) at (5.575,-2.95)
  {\texttt{verif\_forward\_remset.v}
   \nodepart{two}
   \texttt{forward\_remset\_gh}};
\node[proofbox,existing] (vfr) at (5.575,-4.20)
  {\texttt{verif\_forward\_roots.v}
   \nodepart{two}
   \texttt{forward\_roots\_relation}};
\node[proofbox,existing] (vds) at (5.575,-5.45)
  {\texttt{verif\_do\_scan.v}
   \nodepart{two}
   \texttt{do\_scan\_relation}};
\node[proofbox,existing] (vfwd) at (5.575,-6.70)
  {\texttt{verif\_forward.v}
   \nodepart{two}
   \texttt{forward\_relation}};

\draw[prooflink] (gc.east) -- (vgc.west);
\draw[prooflink] (dg.east) -- (vdg.west);
\draw[prooflink] (frs.east) -- (vrs.west);
\draw[prooflink] (fr.east) -- (vfr.west);
\draw[prooflink] (ds.east) -- (vds.west);
\draw[prooflink] (fwd.east) -- (vfwd.west);

\node[specbox,repaired,anchor=north] (specs) at (11.7,0 |- vgc.north)
  {\texttt{GCGraph.v}
   \nodepart{two}
   \texttt{spatial\_gcgraph.v}
   \nodepart{three}
   \texttt{gc\_spec.v}};
\node[theorem,repaired,anchor=north] (iso) at ($(specs.south)+(0,-0.80)$)
  {\texttt{gc\_correct.v}
   \nodepart{two}
   \texttt{garbage\_collect\_isomorphism}};

\draw[prooflink] (vgc.east)
  to[out=0,in=180,looseness=1.18] ($(specs.west)+(0,0.80)$);
\draw[prooflink] (vdg.east)
  to[out=0,in=180,looseness=1.12] ($(specs.west)+(0,0.48)$);
\draw[prooflink] (vrs.east)
  to[out=0,in=180,looseness=1.05] ($(specs.west)+(0,0.16)$);
\draw[prooflink] (vfr.east)
  to[out=0,in=180,looseness=0.78] ($(specs.west)+(0,-0.16)$);
\draw[prooflink] (vds.east)
  to[out=0,in=180,looseness=0.58] ($(specs.west)+(0,-0.48)$);
\draw[prooflink] (vfwd.east)
  to[out=0,in=180,looseness=0.42] ($(specs.west)+(0,-0.80)$);
\draw[prooflink] (specs.south) -- (iso.north);

\begin{scope}[shift={(10.0,-5.95)}]
  \node[legendtext] at (0,0.55) {\textbf{Legend}};
  \draw[call] (0,0.18) -- (0.75,0.18);
  \node[legendtext] at (0.95,0.18) {C call};
  \draw[prooflink] (0,-0.20) -- (0.75,-0.20);
  \node[legendtext] at (0.95,-0.20) {proof/contract link};
  \node[box,existing,minimum width=0.52cm,minimum height=0.26cm,inner sep=0pt] at (0.26,-0.60) {};
  \node[legendtext] at (0.95,-0.60) {pre-existing artifact};
  \node[box,repaired,minimum width=0.52cm,minimum height=0.26cm,inner sep=0pt] at (0.26,-0.98) {};
  \node[legendtext] at (0.95,-0.98) {Codex-assisted repair};
\end{scope}

\end{tikzpicture}}
\caption{Mutable-collector proof map.}
\label{fig:gc-proof-map}
\Description{A three-column diagram showing the C function call graph, the
corresponding VST proof files and abstract effects, the shared model and
specification files, and the final theorem in gc_correct.v.}
\end{figure}

Figure~\ref{fig:gc-proof-map} gives the big-picture map for the repaired
mutable branch and marks the division of work reported here.  The left column
shows the C call graph for the repaired generational collection code, including
the remembered-set path inside \texttt{do\_generation}.  The middle column pairs
each function with the corresponding \VST{} proof file and abstract effect; the
right column shows shared graph and specification files feeding the final theorem
in \texttt{gc\_correct.v}.  Unshaded boxes mark pre-existing artifacts; shaded
boxes mark proof and model artifacts repaired with Codex assistance.

The work began as a conventional human proof effort to strengthen the
implementation, specification, and proof for mutable references.  By
the time Codex was used, the branch already contained substantial
pre-existing artifacts, shown unshaded in
Figure~\ref{fig:gc-proof-map}.  They included lower-level collector
proofs for \texttt{forward}, \texttt{forward\_roots}, and
\texttt{do\_scan}, along with human-written mutable-collector
additions: remembered-set data structures, mutator-side write-barrier
specification work, a refactoring of thread information to carry
remembered-set state, and a verified \texttt{forward\_remset} function
for processing remembered-set entries.  The remaining Codex-assisted
work was to repair the shaded proof and model artifacts and clean them
up.  This phase used Codex as an agentic coding tool: it read the
codebase, edited proof files, ran \texttt{make}, diagnosed errors,
proposed helper lemmas, and performed cleanup.  The goal was not
merely to produce local proof-script fragments, but to state the final
theorem correctly and have its proof accepted by the \Rocq{} kernel.

The Codex-assisted development was organized around three main proof
tasks.  First, the \VST{} proof of \texttt{do\_generation} had to
incorporate the remembered-set \texttt{forward\_remset} function call
and expose the right abstract relation.  Second, the outer
\texttt{garbage\_collect} \VST{} proof had to use that restored
relation while preserving remembered-set invariants across collection
and reset.  Third, \texttt{gc\_correct.v} had to re-establish that the
abstract collection relation still preserves roots-level graph
isomorphism.  The main theorem was accepted before the end of the
effort, but later work still mattered.  That later work audited
assumptions, removed a stale no-backward-edge premise from the
specification path, and simplified the theorem's supporting proof
structure.

The experience was inspired by a recent report
\citep{paraskevopoulou2026machinegenerated} on using Claude Code to
mechanize an ANF transformation correctness proof for CertiCoq.  That
paper showed that an agentic assistant can adapt an existing
compiler-proof technique to a new transformation and produce a
machine-checked proof in a few days.  This report examines a
different point in the design space.  Codex did not start by adapting
a nearby proof architecture whose central assumptions remained valid.
Instead, it worked inside an existing systems-verification development
where mutable updates invalidated the old collector invariant, turning
the task into proof repair rather than proof-architecture adaptation.

The report is intended as a focused experience report rather than a
controlled study.  We contribute the following lessons about
agent-assisted proof maintenance:

\begin{itemize}
\item We document an invariant-repair case in which the central
  problem was not to adapt a related proof architecture, but to
  replace a semantic assumption made false by mutation while
  preserving the final roots-level graph-isomorphism theorem.
\item We describe a concrete division of labor: Codex proposed the
  recorded-backward-edge invariant and helped implement its
  preservation, while we judged that invariant to be the
  right specification for the mutable collector.
\item We report transcript-derived interaction statistics and explain how
those statistics should be interpreted as evidence of supervised, checker-driven
proof maintenance rather than autonomous proof generation.
\item We compare this experience with a recent LLM-assisted compiler-proof
experience report and identify lessons specific to maintaining large
\VST{}/\CertiGraph{} proofs.
\end{itemize}

The claim is not autonomous proof discovery.  Rather, the claim is that agentic
assistance can materially help with checker-driven proof repair when candidate
invariants and specification changes are continuously reviewed by a human proof
engineer.

\section{Background: \CertiGC{}, \VST{}, and Remembered Sets}

\CertiGraph{} \citep{wang2019certigraph} provides a framework for verifying
graph-manipulating C programs: abstract graphs are represented in \Rocq{}, while
separation-logic predicates connect those graphs to concrete heap layouts.
\CertiGC{} is the generational garbage collector in that development.  Its proof
uses \VST{}'s Verifiable C program logic \citep{appel2014programlogics}, whose
soundness theorem connects the logic to CompCert's C semantics
\citep{leroy2009compcert}.
A companion manuscript gives a fuller account of the \CertiGC{} C collector, its
specification, and the underlying \Rocq{} proofs \citep{wang2026certigc}.

The rest of this section supplies the background needed to read
Figure~\ref{fig:gc-proof-map}: how the original collection path differs from
the mutable path, and why that difference forced an invariant change.

Generational collection relies on the observation that most objects die young
\citep[Chapter~9]{jones2023gchandbook}.  The heap is divided into generations
ordered by age; collecting a younger generation copies its live objects to an
older target generation and redirects later uses to the copies.  In
\CertiGC{}, \texttt{garbage\_collect} invokes \texttt{do\_generation} for the
generation currently being collected.  In the original collection path,
\texttt{do\_generation} called \texttt{forward\_roots} and \texttt{do\_scan},
but not \texttt{forward\_remset}.  \texttt{forward\_roots} follows ordinary
program roots, while \texttt{do\_scan} follows fields of already copied objects.
Both ultimately use \texttt{forward}, which copies a referenced live object to
the target generation.

That original path matched the effectively immutable setting that CertiGC first
verified.  Once a graph cell had been allocated, its pointer fields were not
updated, so an older object could not later acquire a pointer to a younger
object.  The proof could therefore rely on the absence of old-to-young edges,
usually stated as a \texttt{no\_backward\_edge} invariant.

Mutable updates, as used for ML-style references or lazy thunk update,
invalidate that invariant.  A program may update a field in an older object to
point to a younger object.  If the collector only follows ordinary roots and
scans copied objects, such a younger object can be missed: the old object that
points to it is outside the younger generation being collected and need not be
scanned.  The mutable branch handles this case by recording old-to-young
pointer writes in a remembered set of source field locations and, as
Figure~\ref{fig:gc-proof-map} shows, by adding \texttt{forward\_remset} to the
\texttt{do\_generation} path before \texttt{forward\_roots} and
\texttt{do\_scan}.  During collection, \texttt{forward\_remset} treats recorded
locations as root-like sources and forwards the younger objects they reach.
With this ordering, subsequent root forwarding and scanning compute the complete
reachability closure from both ordinary program roots and locations recorded in
remembered sets.

For the proof, this program change did more than create new proof-script
obligations.  Keeping the old no-backward-edge invariant would make the proof
easier, but it would verify only heaps that exclude the behavior the mutable
extension is meant to support.  The repair therefore had to change the
specification path itself: the old claim that no old-to-young edges exist had to
be replaced by the weaker recorded-backward-edge invariant, which permits only
recorded old-to-young edges.

A concrete instance fixes the direction of the invariant before the formal
definitions.  Suppose field \texttt{k} of an object \texttt{x} in generation 2
is updated to point to an object \texttt{y} in generation 0.  This creates an
old-to-young edge: \texttt{x} is not part of the generation being collected, but
\texttt{y} may still need to survive that collection.  The write barrier records
the source location \texttt{(x,k)} in generation 0's remembered-set component.
When generation 0 is collected, \texttt{forward\_remset} reads such recorded
locations, follows the pointers stored there, and treats the targets as
additional root-like sources.  The predicate below formalizes this obligation.

At the specification level, this repair keeps the ordinary heap condition and
the remembered-set invariant separate.  The former remains a graph/heap
well-formedness condition that says nothing about remembered-set state:

\begin{lstlisting}
Definition garbage_collect_condition (g: LGraph) (h : part_heap) : Prop :=
  graph_unmarked g /\ no_dangling_dst g /\ ti_size_spec h.
\end{lstlisting}

The mutable-collector obligation is stated by a separate predicate.  It says
that every old-to-young edge is represented in the appropriate remembered set:

\begin{lstlisting}
Definition no_unrecorded_backward_edge (g: LGraph) (rh: remset_heap): Prop :=
  forall e,
    graph_has_e g e ->
    (egeneration e > vgeneration (dst g e))%nat ->
    In (RemSetInterior
          (InteriorVertexPos (fst e) (Z.of_nat (snd e))))
       (nth_remset_space rh (vgeneration (dst g e))).
\end{lstlisting}

In this definition, the comparison
\texttt{egeneration e > vgeneration (dst g e)} selects old-to-young edges: the
edge starts in an older generation and points into a younger one.  The
conclusion then requires the source field position to occur in the
remembered-set space indexed by the destination generation.  Thus the predicate
says that every old-to-young pointer field that may matter to a later collection
has a remembered-set entry for the generation it points into.

The separation between \texttt{garbage\_collect\_condition} and this predicate
makes clear which assumptions concern ordinary graph/heap well-formedness and
which concern mutable-collector remembered-set state.  This predicate is the
recorded-backward-edge invariant used below.  Codex first proposed and
formalized the predicate during the repair session; we accepted it
after semantic review as the right replacement for the old no-backward-edge
condition.

\section{The Proof Task}

The proof task spans two layers of the CertiGC development.  The
\VST{} layer reasons about concrete C states through separation-logic
resources and proves that functions such as \texttt{do\_generation} and
\texttt{garbage\_collect} implement abstract graph relations in
\texttt{GCGraph.v}.  The mathematical layer, especially
\texttt{gc\_correct.v}, reasons about those abstract graphs and proves that
collection preserves the live, roots-reachable part of the graph up to
isomorphism.  The repair had to keep C specifications, abstract relations, and
the graph-isomorphism theorem in sync.

In the remembered-set development, \texttt{rh} denotes the abstract
remembered-set heap threaded through the graph relations.  The variable
\texttt{rmst} denotes remembered-set state used by compatibility predicates that
connect this heap to the graph and to the \VST{}-facing representation
invariants.  This bridge explains why both remembered-set components
appear in the theorem below even though neither appears in its observable
conclusion.

The central proof target was the following theorem.  Its hypotheses expose the
new remembered-set assumptions, while its conclusion retains the original
roots-level graph-isomorphism property:

\begin{lstlisting}
Theorem garbage_collect_isomorphism:
  forall roots1 roots2 g1 h1 rh1 rmst1 g2 h2 rh2 rmst2,
    graph_unmarked g1 -> no_unrecorded_backward_edge g1 rh1 ->
    no_dangling_dst g1 -> roots_graph_compatible roots1 g1 ->
    sound_gc_graph g1 -> remset_graph_state g1 O rmst1 rh1 ->
    remset_heap_covers_graph g1 rh1 ->
    garbage_collect_relation roots1 roots2 g1 h1 rh1 rmst1 g2 h2 rh2 rmst2 ->
    gc_graph_iso g1 roots1 g2 roots2.
\end{lstlisting}

Here the remembered-set heap and state variables are needed to state
collector-private invariants in the hypotheses and relation.  The observable
conclusion, however, is still the ordinary roots-level graph-isomorphism claim.
The two graph arguments \texttt{g1} and \texttt{g2} are the pre- and
post-collection graphs; \texttt{h1}/\texttt{h2} and
\texttt{rh1}/\texttt{rh2} are the corresponding ordinary heaps and
remembered-set heaps, and \texttt{O} is the initial generation index.  This
separation is deliberate: remembered-set bookkeeping is internal collector
state, so exposing it in the user-facing theorem would make the result less
useful.

The invariant flow is as follows.  A mutable update may create an edge from an
older generation to a younger one.  The write barrier records the location of
that edge in the remembered-set space for the destination generation.  During
generation collection, \texttt{forward\_remset} interprets those recorded
locations and forwards the target objects.  Subsequent root forwarding and
scanning then compute the reachability closure from both program roots and
locations recorded in remembered sets.  After the generation is collected and
its remembered-set component is reset, the proof must re-establish that any
remaining backward edge is still recorded in the appropriate remembered set.
Resetting is safe only because the collected generation no longer contains the
live destination objects that would require entries in the reset component;
backward edges into other generations use remembered-set components that are not
reset.
Thus remembered-set entries are treated as additional roots during collection.

The hard part was to bridge remembered-set forwarding, root forwarding,
scanning, and reset into the graph-isomorphism proof.  Forwarding remembered-set
entries temporarily turns recorded old-to-young edges into effective roots.
The one-generation argument proves isomorphism for ordinary roots augmented with
those effective roots.  Because \texttt{gc\_graph\_iso} observes only the
subgraph reachable from its listed roots, the proof then restricts this
augmented-root isomorphism back to ordinary roots.  After that restriction, the
reset proof must show that resetting the collected generation's remembered-set
component does not leave any remaining backward edge unrecorded.  Under the usual graph
and remembered-set compatibility hypotheses, the key preservation lemma has the
following abridged shape.  Codex introduced this form of lemma as the
\VST{}-facing reset-preservation fact once the recorded-backward-edge invariant
was accepted; we reviewed its semantic role and later directed the
core-and-wrapper factoring of related reset-preservation lemmas.

\begin{lstlisting}
Lemma do_generation_relation_no_unrecorded_backward_edge_reset:
  ... ->
  no_unrecorded_backward_edge g rh ->
  remset_compatible g outlier i rmst rh h ->
  do_generation_relation i (S i) roots roots'
    g h rh rmst g_rem h_rem rh' rmst' g' h' ->
  no_unrecorded_backward_edge g' (reset_nth_remset_heap i rh').
\end{lstlisting}

This statement says that one-generation collection may consume the current
generation's remembered set and reset it, but the resulting graph and
remembered-set heap still satisfy the recorded-backward-edge invariant needed by
the outer garbage-collection loop.

The repair therefore required a change in proof architecture, not just patches
to broken proof scripts:

\begin{itemize}
\item remove the invalid global \texttt{no\_backward\_edge} condition;
\item introduce and preserve \texttt{no\_unrecorded\_backward\_edge};
\item make remembered-set entries act as temporary roots during generation
copying;
\item keep the top-level theorem framed as graph isomorphism over ordinary
roots;
\item repair the \VST{} verification files so that the C code and mathematical
relation agree.
\end{itemize}

Table~\ref{tab:proof-architecture} gives the proof architecture used to organize
the repair.  The same invariant appears differently in each proof component: as
spatial resources in \VST{} proofs, as graph facts in the invariant library, as
premises of the abstract collection relation, and finally as assumptions of the
mathematical theorem.  This is why theorem-only repair was insufficient.

\begin{table}[htbp]
\centering
\caption{Proof components and their repair roles.}
\label{tab:proof-architecture}
\scriptsize
\begin{tabular}{@{}L{0.15\linewidth}L{0.25\linewidth}L{0.28\linewidth}L{0.24\linewidth}@{}}
\toprule
Component & Main Files & Contract & Repair Issue \\
\midrule
One-generation \VST{} proof &
\texttt{verif\tus do\tus generation.v} &
The C function implements the generation-step relation after
\texttt{forward\tus remset}, root forwarding, and scanning. &
Reorder the proof around the new call sequence without strengthening scanner
preconditions. \\
Collector-loop \VST{} proof &
\texttt{verif\tus garbage\tus collect.v} &
The loop implements \texttt{garbage\tus collect\tus relation}. &
Preserve remembered-set compatibility and invariants across collection and
reset. \\
Graph invariant library &
\texttt{GCGraph.v}, \texttt{spatial\tus gcgraph.v} &
Backward edges are allowed only when represented in the remembered set. &
Replace global absence of backward edges with preservation of recorded backward
edges. \\
Mathematical theorem &
\texttt{gc\tus correct.v} &
\texttt{garbage\tus collect\tus relation} implies roots-level graph isomorphism. &
Use remembered-set entries as temporary roots while keeping the conclusion about
ordinary roots. \\
Specification boundary &
\texttt{gc\tus spec.v} and \VST{} body lemmas &
\VST{} preconditions justify the assumptions of the mathematical theorem. &
Use an explicit recorded-backward-edge invariant instead of the stale
\texttt{no\tus backward\tus edge} condition. \\
\bottomrule
\end{tabular}
\end{table}

\section{Workflow with Codex}

The Codex-assisted workflow was closer to supervised proof maintenance than to
one-shot proof generation.  The assistant was given access to the repository and
was allowed to edit files and run targeted builds.  We supplied the proof
goal, design constraints, and course corrections.

\paragraph{Proof-component ordering.}
The order of repair followed the proof architecture.  The \VST{} files
first had to establish that the changed C collector still satisfied
the abstract relations used by the mathematical proof.  Only after
\texttt{verif\_do\_generation.v} and
\texttt{verif\_garbage\_collect.v} compiled, establishing that
the C program correctly implements the functional model of the
algorithm, was it productive to repair \texttt{gc\_correct.v}, whose
proof shows that the abstract algorithm preserves roots-level graph
isomorphism.  This avoided treating the final
graph-isomorphism theorem as an isolated \Rocq{} problem disconnected
from the verified C program.

\paragraph{Development loop.}
Most iterations followed a simple pattern.  Codex inspected the current proof
failure, searched for nearby lemmas, edited a focused region, and ran a targeted
build such as:

\begin{lstlisting}
# rebuild definitions and proofs about the graph functional model
make CertiGC/GCGraph.vo
# rebuild proofs that the functional model preserves graph isomorphism
make CertiGC/gc_correct.vo
# rebuild proofs that the C program implements the functional model
make CertiGC/verif_garbage_collect.vo
\end{lstlisting}

Full builds with \texttt{make -j8} were used at integration points.  The
\Rocq{} kernel was the primary feedback mechanism; Codex's natural-language
explanations were useful mainly when the failure indicated a mismatch between
the intended invariant and the proof state.

\paragraph{Proof-state access.}
In addition to shell builds, the session used a locally configured
\texttt{rocq-mcp} Model Context Protocol server
\citep{llm4rocq2026rocqmcp}.  The server exposes Rocq compilation,
environment queries, proof-state startup, tactic checking, multi-tactic
probing, file outlines, and assumption queries as tools for LLM agents.  In
this project it was configured as a repository-local Codex MCP server using the
same \texttt{coqc} binary as the \VST{} build.  Codex could therefore call
tools such as \texttt{rocq\_start}, \texttt{rocq\_check},
\texttt{rocq\_step\_multi}, and \texttt{rocq\_query} from the conversation:
start a proof state at a theorem or error position, inspect the current goals,
query the environment, try small tactic sequences, and then edit the tracked
proof file or run a targeted build.

This made proof-state feedback cheaper and more direct than reading compiler
errors alone.  The development did not include a controlled timing comparison,
so this report does not claim a measured speedup.  Qualitatively, however, MCP
shortened local feedback loops by avoiding many edit-compile-read-error cycles
for exploratory proof-state inspection.  The workflow did not depend logically
on MCP support: similar information can be recovered by compiling focused
temporary files that replay a proof prefix while inserting \texttt{Show}
commands, or by running targeted \texttt{.vo} builds and reading Rocq's error
messages.  MCP was therefore an efficiency and ergonomics tool, not an
assumption behind the proof result.

\paragraph{Invariant proposal and adjudication.}
The recorded-backward-edge invariant was not fixed before the Codex session.  In
the interaction, Codex first formulated the need for a condition informally
phrased as ``no unrecorded backward edge'' and later wrote the first \Rocq{}
definition of that predicate.  We then recognized that the old
\texttt{no\_backward\_edge} premise was semantically invalid for mutable GC and
directed the repair to replace the old invariant rather than keep it as a
convenient hidden assumption.  This episode is representative of the
collaboration: the assistant could propose a useful invariant, but we had
to decide whether accepting it preserved the intended collector specification
rather than merely making the scripts compile.

\paragraph{Human constraints.}
Our guidance was deliberately conservative.  New or changed lemmas were
acceptable, but new or changed \texttt{Definition}s required discussion.
Preconditions were not to be added to \VST{} specs unless there was concrete
evidence that the source program required them.

This constraint appeared early in the scanner proof, where the question was
whether the scanner specification really needed a positive available-space
precondition after remembered-set forwarding.  The accepted direction removed an
overly strong condition; proposed additions such as extra pointer or heap
predicates were rejected unless existing specifications justified them.

\paragraph{Effective uses of Codex.}
Codex was effective at local proof repair and repeated invariant propagation.
Once the recorded-backward-edge invariant and bridge shape were clear, many
subgoals amounted to proving that a property was preserved by a fold over
\texttt{forward\_remset\_item}, by heap updates, or by adding/resetting one
generation.  These tasks are tedious for a human but well suited to an agent
that can search the development, try variants, and run the checker repeatedly.

Codex also helped with cleanup after the theorem was restored.  The branch
contains several later commits whose purpose was not to change the theorem but
to remove stale proof artifacts, factor repeated preservation arguments, audit
the theorem's assumptions against \VST{} specifications, update deprecated
names, and simplify helper lemmas.  In a large proof development, this
maintenance is not cosmetic: stale helper lemmas and stale assumptions obscure
which invariants are actually needed and make future repair harder.

\paragraph{Where human judgment remained central.}
The critical invariant choice was a proof-engineering decision rather than a
tactic choice.  The old proof failed because its semantic assumption was no
longer true.  Although Codex proposed the recorded-backward-edge formulation,
accepting it as the replacement invariant required understanding the collector
algorithm, the write barrier, and the desired external theorem.  Codex could help
explore consequences and fill in preservation lemmas, but the proof had to avoid
shortcuts that would change the result.  Two tempting shortcuts were to
strengthen mathematical specifications with extra heap or remembered-set
preconditions, or to expose remembered-set internals in the main theorem.  Both
changes would have altered the result being verified.  The remembered set is
private collector state; making it part of the theorem would change the public
correctness claim.

\section{Results}

The branch reached a compiling state and was merged into the upstream
\texttt{live} branch as pull request 30 with the merge message ``Replace invalid
\texttt{no\_backward\_edge} invariant for mutable GC'' on May 3, 2026.  The
source repository is \url{https://github.com/CertiGraph/CertiGraph}; the proof
artifact discussed here is \href{https://github.com/CertiGraph/CertiGraph/pull/30}{pull request 30},
whose merge commit is \texttt{85da48e4}.

Table~\ref{tab:repo-evidence} summarizes repository evidence extracted from the
git history.  These numbers should be read as artifact statistics, not as a
precise measure of human or AI labor.  The git author field does not distinguish
manual edits from Codex-assisted edits; the phase boundary below comes from the
transcript-confirmed start of the Codex-assisted work and the corresponding
commit dates.

\begin{table}[htbp]
\centering
\caption{Repository evidence for the mutable-GC proof effort.}
\label{tab:repo-evidence}
\begin{tabular}{lll}
\toprule
Interval & Evidence & Notes \\
\midrule
Nov. 2024--Mar. 26, 2026 & 23 git commits; +5316/-2113 & human groundwork \\
Apr. 22--May 3, 2026 & 51 git commits; +13340/-3557 & Codex-assisted phase \\
Nov. 2024--May 3, 2026 & 74 git commits; +17953/-4967 & whole branch slice \\
\bottomrule
\end{tabular}
\end{table}

Table~\ref{tab:timeline} summarizes the proof-component milestones of this
phase.  The key point is that the main mathematical theorem was
not the last milestone.  \Rocq{} accepted its proof on April 29.  The following
days audited the connection between the \VST{} specification and that theorem,
removed the remaining obsolete invariant premise, and made the proof artifact more
maintainable.

\begin{table}[htbp]
\centering
\caption{Milestone timeline of the Codex-assisted development phase.}
\label{tab:timeline}
\scriptsize
\begin{tabular}{@{}L{0.13\linewidth}L{0.18\linewidth}L{0.39\linewidth}L{0.22\linewidth}@{}}
\toprule
Date & Proof Component & Technical Issue & Outcome \\
\midrule
Apr. 22--26 & One-generation \VST{} proof &
Repair \texttt{verif\tus do\tus generation.v} for the new
\texttt{forward\tus remset}; \texttt{forward\tus roots}; \texttt{do\tus scan}
sequence
without adding unjustified scanner preconditions. &
\texttt{8fc20a3f}: \texttt{verif\tus do\tus generation} compiled. \\
Apr. 26--27 & Collector-loop \VST{} proof &
Repair \texttt{verif\tus garbage\tus collect.v}; preserve remembered-set
compatibility across one-generation collection and reset; separate loop
invariants from function-local specification requirements. &
\texttt{9a235844}: \texttt{verif\tus garbage\tus collect.v} compiled. \\
Apr. 28--29 & Mathematical correctness proof &
Reconstruct \texttt{gc\tus correct.v} around
\texttt{no\tus unrecorded\tus backward\tus edge}, because mutable updates make
the old \texttt{no\tus backward\tus edge} condition false. &
\texttt{94922344}: main theorem proof accepted by \Rocq{}. \\
May 1 & Specification and premise audit &
Check whether the \VST{}-facing \texttt{garbage\tus collect\tus spec} premises justify
the assumptions of the final theorem; remove the stale
\texttt{no\tus backward\tus edge} condition from
\texttt{garbage\tus collect\tus condition}. &
\texttt{51ca1b49}: remembered-set invariant carried explicitly through the \VST{}
specification. \\
May 1--3 & Cleanup and integration &
Factor reset preservation, consolidate remembered-set lemmas, simplify partial-graph
and semi-isomorphism helpers, and rerun full builds after upstream updates. &
PR 30 prepared and merged into \texttt{live}. \\
\bottomrule
\end{tabular}
\end{table}

The May 1 audit began by checking whether the collector specification's
preconditions were sufficient for the assumptions of the final isomorphism
theorem.  That check exposed the stale \path|no_backward_edge| premise path in the
\VST{}-facing proof.  The obsolete condition was not confined to one local
lemma: it flowed from the top-level collector condition into the loop invariant
and then into the preconditions for each generation-collection step.  In code,
this path ran from \path|garbage_collect_condition| through
\path|body_garbage_collect| and \path|gc_cond_implies_do_gen_cons| to
\path|do_generation_condition| obligations.  In other words, commit statistics
alone would have hidden a semantic repair inside what looked like routine proof
maintenance.

This phase touched 18 \CertiGC{} files.  The largest
proof changes were in \texttt{GCGraph.v}, which accumulated the remembered-set
invariants and preservation facts, and \texttt{gc\_correct.v}, which
re-established the graph-isomorphism bridge.  Most verification-facing edits
were in two \VST{} body proofs, \path|verif_do_generation.v| and
\path|verif_garbage_collect.v|.  The collector specification file,
\path|gc_spec.v|, carried the corresponding premise changes.

In addition to repository statistics, interaction statistics were extracted from
the two Codex rollout transcripts corresponding to this work.  These statistics
are meaningful only when their categories are understood as nested rather than
flat.
Table~\ref{tab:transcript-stats} therefore separates additive tool-call counts from
activity tags.  The first tool-call block decomposes the total number of tool
calls, while the activity tags describe overlapping properties of those same
calls.

\begin{table}[htbp]
\centering
\caption{Transcript-derived Codex usage statistics.  The first tool-call block
is additive; activity tags are already included in the tool calls above and are
not mutually exclusive.}
\label{tab:transcript-stats}
\begin{tabular}{@{}L{0.74\linewidth}r@{}}
\toprule
Metric & Value \\
\midrule
\multicolumn{2}{@{}l}{\textit{Interaction records}} \\
\quad Human prompts & 415 \\
\quad Context compactions & 88 \\
\addlinespace[2pt]
\multicolumn{2}{@{}l}{\textit{Tool-call records}} \\
\quad \textbf{Total tool calls} & \textbf{13,305} \\
\qquad Function tool calls & 11,566 \\
\qquad\quad Shell command calls & 8,077 \\
\qquad\quad Shell session inputs & 1,368 \\
\qquad\quad Rocq MCP calls & 2,096 \\
\qquad\quad Other function-tool calls & 25 \\
\qquad Apply-patch edit calls (custom tool calls) & 1,731 \\
\qquad Web search calls & 8 \\
\addlinespace[2pt]
\multicolumn{2}{@{}l}{\textit{Activity tags within the tool calls above}} \\
\quad Proof-checking invocations & 2,131 \\
\quad Shell build commands & 795 \\
\quad Shell search commands & 2,018 \\
\quad Shell read/inspection commands & 5,337 \\
\quad Git commit command invocations & 80 \\
\addlinespace[2pt]
\multicolumn{2}{@{}l}{\textit{Time}} \\
\quad Calendar-span hours & 265.9 \\
\quad Wall-clock hours, per-day sum & 106.5 \\
\quad Codex active hours & 59.3 \\
\bottomrule
\end{tabular}
\end{table}

The parser counts structured rollout items rather than streamed token deltas:
user prompts are \texttt{message} items with role \texttt{user}, excluding
automatic \texttt{<environment\_context>} messages; tool calls are
\texttt{function\_call}, \texttt{custom\_tool\_call}, and
\texttt{web\_search\_call} items; context compactions are top-level
\texttt{compacted} records.  Active time is computed by summing gaps of at most
30 minutes between recorded rollout events, following the same convention as the
related CertiCoq experience report
\citep{paraskevopoulou2026machinegenerated}.  The activity tags in
Table~\ref{tab:transcript-stats} are not additional summands and are not
mutually exclusive: for example, one shell \texttt{make} command can count as a
shell command call, a shell build command, and a proof-checking invocation.  Git
counts are command invocations in the transcripts, not distinct repository
commits, and apply-patch counts are patch tool invocations rather than
semantically distinct edits.  The two rollout files were selected because their
substantive content was the CertiGC mutable-GC repair; no other project logs
were included.  Wall-clock time is reported as the sum of each active day's
first-to-last transcript span; the longer calendar span is included only to make
clear that the transcript spans multiple calendar days.  The statistics end at
the mutable-collector cutoff date, May 3, 2026; later transcript entries about
writing this report are excluded.

These numbers characterize a supervised, checker-driven workflow.  The 59.3
Codex active hours and 415 human prompts show sustained interaction rather than
a single autonomous generation attempt.  The large number of shell, search,
read, and build commands reflects the main mode of work: Codex repeatedly
navigated the existing repository, made small proof edits, and used the \Rocq{}
kernel to test whether each local repair matched the surrounding proof
architecture.

The aggregate statistics hide an important distinction between theorem repair
and post-theorem cleanup, so Table~\ref{tab:per-day-stats} breaks the same
categories down by day.

\begin{table}[htbp]
\centering
\caption{Per-day Codex development activity.  \textsc{Tools} is the total;
\textsc{Shell}, \textsc{Inputs}, \textsc{Rocq}, and \textsc{Edits} are major
subcounts, while \textsc{Checks}, \textsc{Build}, \textsc{Search},
\textsc{Read}, and \textsc{Git} are activity tags.}
\label{tab:per-day-stats}
\footnotesize
\setlength{\tabcolsep}{2pt}
\begin{tabular}{@{}l*{14}{r}@{}}
\toprule
Day & Prompts & Compact & Tools & Shell & Inputs & Rocq & Checks & Build & Search & Read & Edits & Git & Wall(h) & Active(h) \\
\midrule
Apr. 22 & 24 & 2 & 539 & 384 & 74 & 20 & 67 & 60 & 111 & 210 & 54 & 4 & 5.7 & 2.9 \\
Apr. 23 & 2 & 0 & 52 & 32 & 1 & 18 & 17 & 1 & 11 & 20 & 1 & 0 & 0.8 & 0.8 \\
Apr. 26 & 42 & 8 & 1,750 & 912 & 346 & 255 & 303 & 212 & 195 & 482 & 234 & 10 & 9.0 & 7.6 \\
Apr. 27 & 58 & 15 & 2,499 & 1,326 & 417 & 503 & 474 & 168 & 342 & 878 & 250 & 20 & 14.1 & 10.1 \\
Apr. 28 & 93 & 26 & 3,070 & 1,837 & 32 & 727 & 602 & 82 & 516 & 1,296 & 465 & 7 & 23.0 & 12.8 \\
Apr. 29 & 62 & 23 & 2,705 & 1,657 & 118 & 546 & 402 & 29 & 440 & 1,215 & 381 & 11 & 21.4 & 10.8 \\
Apr. 30 & 53 & 4 & 759 & 536 & 129 & 4 & 61 & 57 & 102 & 311 & 82 & 9 & 11.6 & 4.9 \\
May 1 & 43 & 6 & 1,031 & 686 & 211 & 0 & 110 & 110 & 153 & 447 & 134 & 5 & 10.8 & 4.9 \\
May 2 & 20 & 2 & 487 & 401 & 2 & 12 & 60 & 52 & 79 & 262 & 72 & 6 & 5.3 & 2.1 \\
May 3 & 18 & 2 & 413 & 306 & 38 & 11 & 35 & 24 & 69 & 216 & 58 & 8 & 4.8 & 2.3 \\
\midrule
Total & 415 & 88 & 13,305 & 8,077 & 1,368 & 2,096 & 2,131 & 795 & 2,018 & 5,337 & 1,731 & 80 & 106.5 & 59.3 \\
\bottomrule
\end{tabular}
\end{table}

The daily breakdown makes the artifact story visible in the interaction data.
The largest activity peak occurs on Apr. 28--29, with 5,775 tool calls and
1,273 Rocq MCP calls, matching the reconstruction of \texttt{gc\_correct.v}.
The May 1--3 tail is smaller but still substantial: 81 prompts, 1,931 tool
calls, and 19 git command invocations after the main theorem proof had already
been accepted.  This supports the claim that theorem acceptance was followed by
substantive premise audit, cleanup, and integration work rather than by a final
build alone.

The transcript-derived file activity aligns with the git history.  The files
with the most patch calls were \texttt{gc\tus correct.v} (836),
\texttt{GCGraph.v} (384), \texttt{verif\tus garbage\tus collect.v} (226), and
\texttt{verif\tus do\tus generation.v} (131).  The most frequent targeted shell
builds were \texttt{verif\tus garbage\tus collect} (208), \texttt{GCGraph}
(163), \texttt{verif\tus do\tus generation} (122), and \texttt{gc\tus correct}
(119).  The proof artifact has four important final properties.  The C collector
calls \texttt{forward\tus remset} before root forwarding and scanning; the
tracked proof sources no longer rely on the invalid
\texttt{no\tus backward\tus edge} invariant; the \VST{}-facing collector
specification carries the recorded-backward-edge invariant explicitly while the
ordinary graph/heap condition remains independent of remembered-set state; and
\texttt{garbage\tus collect\tus isomorphism} keeps the observable conclusion at
ordinary roots-level graph isomorphism.  Remembered-set assumptions remain
proof-side obligations, not part of that observable theorem.

\section{Observations}

\paragraph{The theorem statement is the artifact to defend.}
In a large formal development, a compiling proof is necessary but not sufficient.
In this case, that defense meant guarding against specification drift:
strengthening preconditions until the proof became easier, or letting
remembered-set details leak into a theorem whose purpose was to state graph
preservation for ordinary roots.  Definitions and preconditions are part of the
scientific claim, not local proof conveniences; changing them required evidence
that the collector semantics needed the change.  Human review therefore focused
on definitions, assumptions, and theorem statements.

\paragraph{LLMs are useful for local invariant propagation.}
Once a good invariant is chosen, a large amount of proof work is mechanical but
not trivial.  The proof assistant requires exact lemmas about folds, list
updates, graph validity, generation bounds, heap sizes, and compatibility
predicates.  Codex was useful because it could navigate the existing library and
produce these bridge lemmas incrementally.  The resulting proof still benefited
from cleanup: some helper lemmas were introduced only to cross a local gap and
became unnecessary after the proof architecture stabilized.

\paragraph{Existing libraries are both leverage and constraint.}
\VST{} and \CertiGraph{} made the result possible because they already provided
the semantic model, spatial graph representation, and verification discipline.
They also constrained the shape of the repair.  A small change to a definition
could ripple through many files, and automation failures were often symptoms of
larger mismatches between mathematical and spatial invariants.  Codex was more
effective when the development stayed within existing patterns than when it
required new proof architecture.

\paragraph{A failed invariant is a result.}
An important outcome of this phase was that the old
no-backward-edge strategy was not a proof obligation waiting to be discharged.
It was the wrong condition for mutable collection.  Treating that failure as
information changed the task from local script repair to invariant replacement.
This is a useful distinction in LLM-assisted verification because an assistant
can both propose a better invariant and make local progress under an
overly strong assumption.  Human review is needed to identify the path that
preserves the intended theorem.

\paragraph{A completed machine-checked proof is not the end of validation.}
The branch did not stop when \Rocq{} first accepted the top-level theorem
\texttt{garbage\_collect\_isomorphism}.  The remaining question was whether the
\VST{} specification path justified exactly the assumptions of that theorem.
The May 1 premise audit removed the obsolete
no-backward-edge path from \VST{}-facing conditions and made the
recorded-backward-edge condition the explicit mutable-GC invariant.
Subsequent cleanup factored repeated reset-preservation arguments, simplified
helper lemmas, and updated deprecated \Rocq{} names.  In a mechanized
development, this cleanup also exposes the proof's real dependencies.

\section{Comparison with Related Accounts}

The closest inspiration is the CertiCoq report on machine-generated compiler
proofs.  That report \citep{paraskevopoulou2026machinegenerated} describes
using Claude Code to mechanize the ANF transformation correctness proof.  That work
used the existing CPS proof as a template and reported approximately 7,800 lines
of \Rocq{} proof developed in about 96 hours.  The human authors did not write
proof scripts directly; they supplied high-level guidance and reviewed the
result.

The present experience is similar in three respects.  First, both tasks are
substantial mechanized proofs in established proof-assistant verification
infrastructure.
Second, both rely on the proof assistant as a machine checker, not on the
language model's self-assessment.  Third, both show that an agent can make
progress when it can repeatedly search a mature codebase, edit proof files, and
invoke the checker in the same loop.

The differences are more instructive.  The ANF report is primarily an
adaptation story: a closely related CPS proof supplied a proof architecture that
Claude Code could adapt to ANF, although the new proof still required genuinely
new reasoning.  The mutable-\CertiGC{} work is an invariant-repair and
proof-maintenance story.  Its nearby proof path was not a reusable architecture,
because mutable updates made the central mathematical assumption false.  The
central problem was to choose a new invariant and thread it through existing
\VST{}, \CertiGraph{}, and collector-correctness layers without weakening the
final theorem.

A second difference is that the first compiling version of the main theorem was
not the endpoint.  The post-theorem premise audit changed the
\VST{}-facing specification path by removing a stale immutable-collector
assumption.  That phase is less visible if one measures only theorem completion,
but it was essential to making the final artifact match the mutable-collector
semantics.

This distinction affects how one should evaluate the assistant.  When a related
proof architecture is available, productivity can be measured by how quickly the
assistant adapts it while filling in the genuinely new reasoning.  For invariant
repair, the main question is whether the assistant helps converge on the right
architecture while preserving the semantic contract.  In this case, Codex was
more like a persistent proof maintenance collaborator than an autonomous prover.

A contemporaneous non-archival report \citep{sergey2026moveborrow} describes an
AI-assisted Lean mechanization of the Move borrow checker's type soundness
theorem.  Although that work is PL metatheory rather than \VST{} program
verification, it reports a similar pattern: the assistant handled preservation
cases, invariant propagation, and proof repair, while the human researcher
supplied the decisive semantic structure.  Together, these reports suggest that
agentic assistants work best with a machine checker validating small edits and
human review guarding semantic structure.

\section{Limitations and Validation}

This is a single-case experience report, not a controlled experiment.  The work
was performed on a branch that already contained substantial human design and
proof effort.  The repository history provides dates, commits, and diffs, but it
does not by itself identify which lines were written manually and which were
written with Codex.  The quantitative transcript statistics reported above are
retrospective, and the full rollout files contain private project discussion.
They are therefore treated as local evidence rather than public replication
artifacts.

The strongest validation is machine checking.  The relevant proof files compile,
and the branch was merged upstream.  Nevertheless, the checker validates only
the statements that remain in the code.  Human review is still required to judge
whether those statements are the intended ones.

The stale \texttt{no\_backward\_edge} assumption was not introduced during the
Codex-assisted phase.  It came from the earlier proof of the effectively
immutable collector and became invalid only when mutable updates entered the
collector model.

This account also does not claim that Codex independently discovered the whole
remembered-set proof architecture.  Codex did propose the
\texttt{no\_unrecorded\_backward\_edge} formulation and wrote its first \Rocq{}
definition, but accepting that formulation as the replacement for
\texttt{no\_backward\_edge} required our semantic review.  Our role also
included understanding the collector semantics, constraining acceptable
specification changes, and deciding when the proof had to reject an old
assumption rather than preserve it.  The more defensible claim is that an
agentic assistant substantially helped turn that proof-engineering direction
into a compiling, cleaned-up artifact.

The interaction also exposed tooling limits.  Codex mostly consumed the same
textual build errors and proof states a human would see.  More structured
machine-readable feedback from \Rocq{} and from \VST{} proof obligations could
make this style of repair less dependent on repeated shell builds and local
search.  The experience suggests that proof-assistant support for agentic tools
should expose not only errors, but also obligation structure, dependency
information, and stable references to nearby applicable lemmas.

Full proof builds during the Codex-assisted development phase used
\texttt{make -j8} in the \texttt{certigraph} opam switch with \Rocq{} 9.1.1, OCaml 4.14.3,
CompCert 3.17, and \VST{} 2.16.  Because this report does not report wall-clock
proof-build duration, it omits hardware-dependent build-time measurements.

\section{Conclusion}

The mutable-\CertiGC{} branch shows a pragmatic mode of AI-assisted formal proof
development.  Codex did not replace the need for proof-engineering judgment, but
it was useful once candidate invariants were judged against the collector
semantics and we kept specification boundaries explicit.  The
resulting proof completes the mutable-collector story by replacing an invalid
no-backward-edge assumption with the recorded-backward-edge invariant while
preserving graph isomorphism over ordinary program roots rather than collector
bookkeeping.

The broader lesson is that agentic assistants can already help with large
machine-checked developments when the collaboration is organized around the
checker and around carefully guarded theorem statements.  They are especially
valuable for local repair, invariant propagation, and cleanup in mature
codebases.  The hard work remains deciding what should be true.

\begin{acks}
The author thanks Andrew Appel for suggesting this experience report and for
discussions about \CertiGC{} and AI-assisted proof development.  The author also
thanks Zoe Paraskevopoulou for making her related CertiCoq experience report
available; its successful use of Claude Code directly motivated this
agent-assisted \CertiGC{} proof effort.
\end{acks}

\bibliographystyle{ACM-Reference-Format}
\bibliography{references}

\end{document}